\documentclass[]{aastex631}

\usepackage{amsmath}

\begin{document}

\title{Seafloor Weathering and Stochastic Outgassing Unlikely to Significantly\\ Shorten the Future Lifespan of Earth's Terrestrial Biosphere}

\author{Livia Zhu}
\affiliation{Department of Geophysical Sciences, University of Chicago\\            
5734 S. Ellis Ave.\\ 
Chicago, IL 60637, USA}

\author[0000-0001-9289-4416]{R.J. Graham}
\affiliation{Department of Geophysical Sciences, University of Chicago\\            
5734 S. Ellis Ave.\\ 
Chicago, IL 60637, USA}

\author[0000-0001-8335-6560]{Dorian S. Abbot}
\affiliation{Department of Geophysical Sciences, University of Chicago\\            
5734 S. Ellis Ave.\\ 
Chicago, IL 60637, USA}
\correspondingauthor{Livia Zhu}
\email{ljz@uchicago.edu}



\begin{abstract}
Current understanding suggests that as the Sun brightens in the far future, Earth’s carbonate-silicate cycle will offset increasing temperatures by drawing CO$_2$ out of the atmosphere, ultimately leading to the extinction of all terrestrial plant life via either overheating or CO$_2$ starvation. Most previous estimates put the future lifespan of Earth's terrestrial biosphere at $\sim$1 billion years, but recent work used a new coupled climate-continental weathering model with up-to-date parameter constraints to revise this estimate upward to 1.6-1.86 billion years. In this study, we extend the model to examine the impacts of seafloor weathering and stochastic variations in CO$_2$ outgassing rates on the remaining lifespan of Earth’s terrestrial biosphere. We find that if seafloor weathering has a stronger feedback than continental weathering and accounts for a large portion of global silicate weathering, then the remaining lifespan of the terrestrial biosphere can be shortened, but a lifespan of more than 1 billion years (Gyr) remains likely. Similarly, stochastic fluctuations in outgassing rates can have a significant impact if the size of the fluctuations exceed those observed over the last 1 billion years. The impact of weak seafloor weathering and lower variability stochasticity are minor. Our work provisionally supports a lengthened lifespan of Earth's terrestrial biosphere, suggests robustness of this lengthened lifespan to planetary parameters that may vary among exoplanets, and identifies seafloor weathering as a key process that requires further study and constraint.
\end{abstract}



\section{Introduction}\label{sec:intro}

The Sun's luminosity is slowly increasing, which will significantly impact life's ability to survive at Earth’s surface in the distant future \citep[e.g.,][]{caldeira1992life}. As Earth's surface warms due to increasing insolation, the silicate weathering feedback should partially offset rising temperatures by reducing the concentration of CO$_2$ in the atmosphere \citep{Walker-Hays-Kasting-1981:negative}, continuing a process that has maintained favorable conditions for life on Earth for billions of years \citep{franck1999modelling}. Eventually, however, Earth's terrestrial biosphere will come to an end. The silicate weathering feedback could be overwhelmed by the brightening Sun, causing temperatures to rise above the overheating threshold for plant life \citep[e.g.][]{lenton2001biotic, franck2006causes}, or, ironically, the feedback could be so effective that it drives CO$_2$ concentrations below the minimum threshold necessary for photosynthesis \citep{lovelock1982life, caldeira1992life}. Either scenario would result in the extinction of all terrestrial plants and the animals that depend on them.

Most work in this area has estimated a remaining lifespan of the terrestrial biosphere of approximately 1 billion years \citep[e.g.,][]{caldeira1992life, lenton2001biotic, franck2002long, VonBloh2003, franck2006causes, rushby2018long, SousaMello2020, ozaki2021future, mello2023planetary} with extinction caused by CO$_2$ starvation. Recently, \citet{graham2024substantial} suggested that this lifespan may be significantly longer because previous studies used silicate weathering parameters that are inconsistent with current best estimates \citep{krissansen2017constraining} and assumed overly conservative overheating and CO$_2$ starvation thresholds for land plants. With the adjustments made in \citet{graham2024substantial}, the remaining lifespan of the terrestrial biosphere could extend as far as 1.86 Gyr, and overheating reemerges as the likely kill mechanism.

\citet{graham2024substantial} made a number of simplifying assumptions that could affect their conclusions. For example, they only considered continental silicate weathering. ``Seafloor weathering,'' the dissolution of basaltic silicates on the ocean floor, could also produce a negative feedback stabilizing Earth's climate, potentially with a stronger temperature dependence than its continental counterpart \citep{Brady:1997p3530, coogan2013evidence, coogan2015alteration, krissansen2017constraining, mello2023planetary}. Seafloor weathering could therefore reduce the equilibrium atmospheric CO$_2$ in the distant future, which would shorten the lifespan of the terrestrial biosphere in CO$_2$ starvation scenarios. Another potentially important factor is variation in CO$_2$ outgassing rates throughout Earth history \citep{wordsworth2021likely, baum2022simple}. The addition of CO$_2$ to the atmosphere via volcanic activity is modulated by Earth's tectonic processes \citep{franck2000reduction}. Natural variation in plate movement and volcanism have contributed to historical variations in outgassing \citep{mills2021spatial}, and they will continue to do so in the far future. Randomly occurring periods of exceptionally low or high outgassing may cause major stress to Earth's plant life, cutting the future lifespan short with early overheating or CO$_2$ starvation. Most previous work on the lifespan of the terrestrial biosphere does not include seafloor weathering, and no study considers stochastic outgassing.

In this study, we extend the modeling framework from \citet{graham2024substantial} to include seafloor weathering as well as allow for variation in CO$_2$ outgassing rates. With these modifications, we then re-evaluate the future lifespan of the terrestrial biosphere. We find that the impact of seafloor weathering varies drastically with modern seafloor weathering fraction and feedback strength. If seafloor weathering is a very strong negative feedback, then the upper limit to the remaining lifespan of the terrestrial biosphere could be shortened, although a lifespan of at least 1 billion years remains likely. Similarly, the impact of stochastic outgassing changes considerably with the size of variability. If the size of the stochastic fluctuations meet or exceed the variability observed over the last 1 billion years, then there is a significant probability of early biosphere death before 1 billion years due to advance overheating or CO$_2$ starvation. In contrast, the impacts of lower stochastic variability and weak seafloor feedback are minor. 

\section{Modeling and Methods}

We use the coupled climate-weathering modeling framework of \citet{graham2024substantial} with the modifications described below.

\subsection{Weathering Model}
We use a standard silicate weathering model with power law CO$_2$ dependencies and exponential temperature dependencies for continental ($W_\mathrm{L}$) and seafloor ($W_\mathrm{S}$) weathering components of the carbonate-silicate cycle \citep{Walker-Hays-Kasting-1981:negative, abbot2016analytical, krissansen2017constraining, chambers2020effect, mello2023planetary, graham2024substantial}:
\begin{align}\label{eqn:weathering}
W_\mathrm{L} &= {W_{\mathrm{0, L}}}(\frac{\mathrm{CO}_\mathrm{2,soil}}{\mathrm{CO}_\mathrm{2,soil,0}})^{\beta_\mathrm{L}} \times \mathrm{exp}({\frac{T-T_\mathrm{0}}{T_{\mathrm{e,L}}}})\\
W_\mathrm{S} &= {W_{\mathrm{0,S}}}(\frac{\mathrm{CO}_{\mathrm{2,atm}}}{\mathrm{CO}_{\mathrm{2,atm,0}}})^{\beta_\mathrm{S}} \times \mathrm{exp}({\frac{T_\mathrm{pore}-T_{\mathrm{pore},0}}{T_\mathrm{e, S}}})
\end{align}
where $W_{0,L}$ represents the modern rate of weathering on the continents (L); $T$ is Earth's global-mean surface temperature; $T_\mathrm{0}$ is Earth's pre-industrial global-mean surface temperature, 288 K. ${T\mathrm{_{e,L}}}=31$ and $\beta_{\mathrm{L}}=0.41$ represent the temperature and CO$_2$ dependencies of continental weathering respectively. $\mathrm{CO}_\mathrm{2,soil}$ is the soil CO$_2$, calculated as a function of atmospheric CO$_2$ ($\mathrm{CO}_\mathrm{2,atm}$) and plant productivity. See \citet{graham2024substantial} for full descriptions of the plant productivity and soil CO$_2$ models used in this study. 
${T\mathrm{_{e,S}}}$ and $\beta_{\mathrm{S}}$ represent the temperature and CO$_2$ dependencies of seafloor weathering respectively, the latter of which can be converted from a pH dependency as in \citet{krissansen2018constraining,hayworth2020waterworlds}: $$\label{eqn:ph} 
\beta_{\mathrm{S}} = \frac{\gamma}{1.34}$$ 
where $\gamma$ is the pH dependence ranging from 0.04 to 0.48 with a median value of 0.27, producing a corresponding $\beta_{\mathrm{S}}$ CO$_2$ dependence ranging from 0.03 to 0.36 with median 0.20. 

Following \citet{krissansen2017constraining}, we assume the temperature that seafloor silicates are exposed to, $T_\mathrm{pore}$, is linearly related to surface temperature, $T_\mathrm{S}$. $$\label{eqn:pore_temp}
T_{\mathrm{pore}} = a_{\mathrm{grad}} \times T_\mathrm{S} + b_{\mathrm{int}},$$
where $a_\mathrm{grad}=1.02$ and $b_\mathrm{int}=-7.66$~K. Here $b_\mathrm{int}$ is derived from Equation 12 in \citet{krissansen2017constraining} with their additional conversion constant of 9~K to account for the difference between deep sea and pore temperatures. 

On long timescales weathering balances outgassing ($V = W_\mathrm{L} + W_\mathrm{S}$), and on modern Earth $V_0 = W_\mathrm{0,L} + W_\mathrm{0,S}$, where $V_0$ is the modern rate of CO$_2$ outgassing, approximately 7 Tmol yr$^{-1}$ \citep[e.g.,][]{catling2017atmospheric}. We define $\alpha = \frac{W_{\mathrm{0, S}}}{W_{\mathrm{0, L}}}$ so that
\begin{align}\label{eqn:weathering defs balance}
W_\mathrm{0,S} &=\frac{\alpha V_0}{1+\alpha}, \\ 
W_\mathrm{0,L} &= \frac{V_0}{1+\alpha}.  
\end{align}

To investigate the impact of seafloor weathering, we calculate the future lifespan of the terrestrial biosphere as we vary T$_\mathrm{e,S}$, $\beta_{S}$, and $\alpha$, all of which are poorly constrained. We choose T$_\mathrm{e,S}$ to range from 6.5 to 62 K, and $\beta_{S}$ to range from 0.03 to 0.36. $\alpha$ ranges from 0.027 to 0.184. The parameter ranges we consider are the 90\% confidence interval ranges estimated from a fit to paleoclimate proxy data \citep{krissansen2017constraining}, with the exception of the upper bound of T$_\mathrm{e,S}$, which is poorly constrained by the posterior distributions of \citet{krissansen2017constraining}. Instead, we consider values of T$_\mathrm{e,S}$ up to twice the median T$_\mathrm{e,L}$, at which point seafloor weathering has minimal impact on our results.

\subsection{Outgassing Model} 

\begin{figure}[htb]
    \centering
    \textbf{OU Model Stochastic Outgassing vs. Outgassing Reconstructions}
    \includegraphics[width=0.9\linewidth]{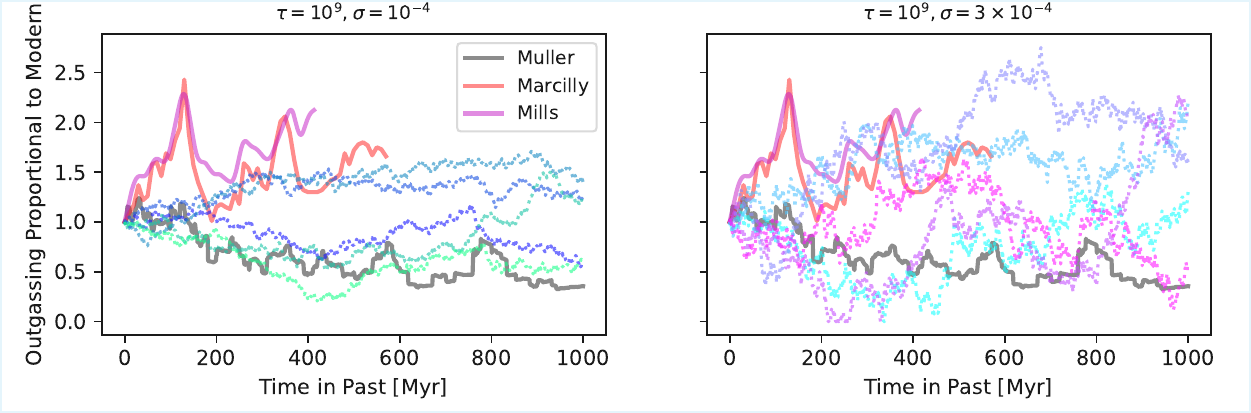}
    \caption{\textbf{The OU stochastic outgassing model output is qualitatively consistent with outgassing reconstructions within the last 1 billion years.} The 5 dotted lines in each subplot represent instances of our model output, and the three solid lines represent the reconstructions of \citet{mills2021spatial}, \citet{marcilly2021new}, and \citet{muller2024degassing}. The parameters used for the model are $\tau = 10^{9}$ and $\sigma = 10^{-4}$ on the left hand side, and $\tau = 10^{9}$ and $\sigma = 3 \times 10^{-4}$ on the right hand side. The relative concentration of the simulated outgassing patterns between the reconstructions of \citet{mills2021spatial} and \citet{muller2024degassing} on the left suggests that larger variability is likely. The right subplot examines variability more consistent with the reconstructions.}
    \label{fig:modelcompare}
\end{figure}

Following \citet{baum2022simple}, we model stochastic outgassing using an Ornstein-Uhlenbeck process,
\begin{align}\label{eqn:outgassing}
dV = \frac{1}{\tau}(\mu_\mathrm{t} - V)dt + \sigma dB
\end{align}
where $\mu_\mathrm{t}$ is the mean outgassing rate at time $t$; $\tau$ is the mean-reversion time; $\sigma$ is the noise magnitude; and $dB$ denotes the Wiener process. The outgassing and the weathering model parameters are chosen independently. We create a wide ensemble of stochastic simulations with varying values of $\tau$ and $\sigma$. Outgassing proxies over the last 720 million years qualitatively suggest that $\tau$ is larger than 100 million years \citep{mckenzievolcanism2016,baum2022simple}, and \citet{baum2022simple} found that models with smaller $\tau$ exhibit a preference for snowball events later in Earth's history, contradicting their absence over the Phanerozoic. So, based on the results of \citet{baum2022simple}, we let $\tau$ range from $10^8$ - $10^9$ years. Similarly, we choose $\sigma$ values that \citet{baum2022simple} find to generally maintain temperatures above Snowball levels across Earth history without eliminating the possibility of global glaciation entirely, $\sigma = 10^{-5}$ - $10^{-3}$ Tmol year$^{-1}$. These ranges of $\sigma$ and $\tau$ contain values that produce outgassing simulations visually consistent with the outgassing reconstructions of \citet{mills2021spatial}, \citet{marcilly2021new}, and \citet{muller2024degassing} from the last 400 million to 1 billion years (Fig. \ref{fig:modelcompare}). 

The stochastic regime represented in the right subplot better captures the variability present in the reconstructions. We therefore use $\sigma = 3 \times 10^{-4}$ to mark the threshold between low and high variability, with high variability encompassing $\sigma >3 \times 10^{-4}$ and corresponding to stochastic variability larger than that of the three reconstructions examined here.

\section{Results}

\begin{figure}[ht]
    \centering
    \textbf{Surface Temperature, Atmospheric CO$_2$, and Weathering Rate Time Series}
    \includegraphics[width=0.95\linewidth]{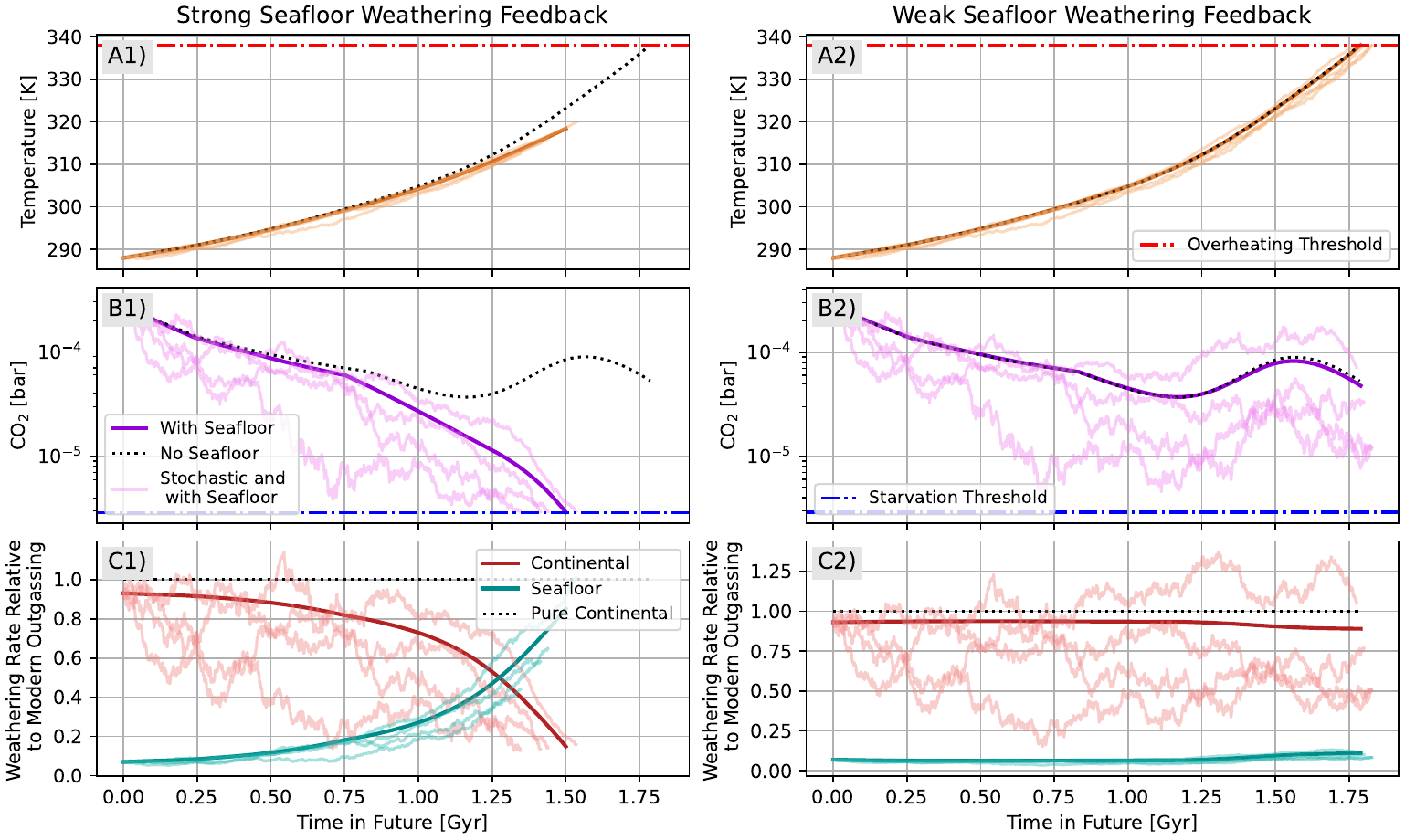}
    \caption{\textbf{Strong seafloor weathering shortens the lifespan of the terrestrial biosphere by nearly 300 million years and shifts the biosphere kill mechanism from overheating to CO$_2$ starvation, whereas weak seafloor weathering has little effect. Stochastic CO$_2$ outgassing does not significantly impact results.} Time series of surface temperature (A), atmospheric CO$_2$ (B), and weathering rates (C) with a strong seafloor weathering feedback ($T_\mathrm{e,S} = 9.07\ K < T_\mathrm{e,L}$, left column) and a weak seafloor weathering feedback ($T_\mathrm{e,S} = 62\ K > T_\mathrm{e,L}$, right column). Results without seafloor weathering (dotted lines), without stochastic CO$_2$ outgassing (dark colored lines), and for different realizations of stochastic CO$_2$ outgassing (faint colored lines) are shown in each panel. Other parameter values are $\tau = 10^9$, $\sigma = 10^{-4}$, $\beta_\mathrm{S} = 0.20$, $\alpha$ = 0.075,  $\beta_\mathrm{L}=0.41$, and $T_\mathrm{e,L}=31$~K.}
    \label{fig:timestack}
\end{figure}

The death of the complex terrestrial biosphere is caused either by overheating or CO$_2$ starvation of plant life, and therefore the time to death is determined by a combination of surface temperatures and atmospheric CO$_2$ concentrations. With default parameters and in the absence of seafloor weathering, atmospheric CO$_2$ concentrations stay above the starvation threshold for the next 1.86 billion years, and the extinction of the terrestrial biosphere is caused by overheating \citep[Fig.~\ref{fig:timestack},][]{graham2024substantial}. Results are similar with a weak seafloor weathering feedback (e.g. $T_\mathrm{e,S} = 62\ K > T_\mathrm{e,L} = 31\ K$, Fig.~\ref{fig:timestack}). When the seafloor weathering feedback is strong (e.g. $T_\mathrm{e,S} = 9.07\ K < T_\mathrm{e,L} = 31\ K$), on the other hand, the lifespan of the complex terrestrial biosphere is reduced by approximately 300 million years (Fig.~\ref{fig:timestack}). In this case the distinctive reversal of the decline in CO$_2$ due to reduced plant productivity \citep{graham2024substantial} disappears, and CO$_2$ declines monotonically. As a result, the kill mechanism switches from overheating to CO$_2$ starvation. 

Stochastic CO$_2$ outgassing does not change the overall trends discussed above, though the impact of stochastic outgassing on atmospheric CO$_2$ through time is much larger than its impact on surface temperature, suggesting that it will have a larger effect on the lifespan if the kill mechanism is CO$_2$ starvation, as in the case of strong seafloor weathering above. Notably, we see instances where stochasticity both extends and shortens the future lifespan of the terrestrial biosphere. These extensions and shortenings, however, are minor for our choice of noise magnitude and mean-reversion time in Figure \ref{fig:timestack} (less than 200 million years). 

\begin{figure}[ht]
    \centering
    \textbf{Fraction of Simulations Terminated Under Six Stochastic Outgassing Regimes}
    \includegraphics[width=0.9\linewidth]{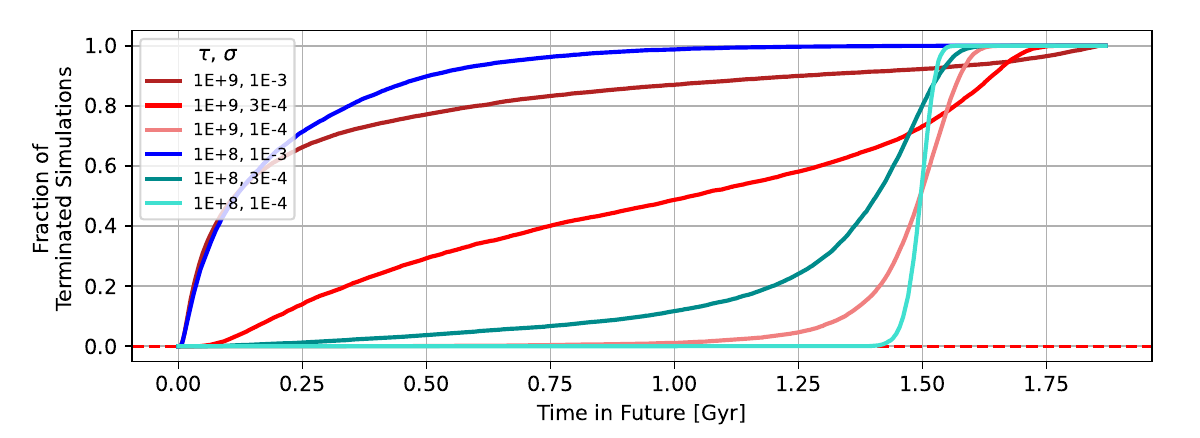}
    \caption{\textbf{Stochastic CO$_2$ outgassing has a major effect on the future lifespan of the terrestrial biosphere when $\sigma \ge 3\times10^{-4}$ Tmol Year$^{-1}$.} CDF displaying the proportion of simulations that have reached biosphere death over time. Six combinations of $\tau$ and $\sigma$ are explored, each featuring ensembles of realizations of stochastic CO$_2$ outgassing of size N=10000. With no stochasticity, the expected future lifespan is 1.50 billion years. Other parameter values are $\beta_\mathrm{S} = 0.20$, $T_\mathrm{e,S}=9.07$~K, $\alpha$ = 0.075, $\beta_\mathrm{L}=0.41$, and $T_\mathrm{e,L}=31$~K. This corresponds to the strong seafloor weathering feedback regime that leads to the CO$_2$ starvation kill mechanism.}
    \label{fig:Fractiondead}
\end{figure}

When we consider large ensembles of stochastic CO$_2$ outgassing realizations and the full range of plausible noise magnitudes and mean-reversion times (Fig.~\ref{fig:Fractiondead}), we observe that the impact of stochastic outgassing is asymmetric. Stochasticity has the potential to shorten the future lifespan of the terrestrial biosphere more significantly than it does to lengthen it. In Figure \ref{fig:modelcompare}, the stochastic regime represented in the right subplot corresponds to the bright red line in Figure \ref{fig:Fractiondead}. While it captures well the variability present in the three reconstructions, it also corresponds to a sizable proportion of early biosphere deaths. With this amount of variability, outgassing often dips below the threshold necessary to induce CO$_2$ starvation long before the expected time of overheating death in the absence of stochasticity. This phenomenon, observed explicitly in Figure \ref{fig:Fractiondead}, is even more dramatic when $\sigma = 10^{-3}$. These highly variable outgassing patterns frequently cause atmospheric CO$_2$ levels to dip below the CO$_2$ starvation threshold. When $\sigma = 10^{-4}$, the future lifespans are rarely ever shortened below 0.8 Gyr (0.58\%). For $\sigma = 10^{-3}$, on the other hand, the majority of simulations have reached biosphere death by 0.8 Gyr (84.18\%). For high variability in the range of $3\times10^{-4}$ to $10^{-3}$ Tmol Year$^{-1}$, the OU model often produces near-zero outgassing values, and future modeling of stochastic outgassing may choose to model log outgassing or use other means in order to preclude common instances of zero or near-zero outgassing rates.

We also note that as Earth's mantle cools, tectonic activity will slow, leading to a reduction in volcanic activity and the CO$_2$ outgassing rate \citep{Franck:1999p2360, chambers2020effect}. To investigate the effect of this process, we also considered an alternative outgassing scheme in which the mean outgassing rate, $\mu_\mathrm{t}$, decays with time following \citet{chambers2020effect}:
\begin{align}\label{eqn:meandecay}
\mu_\mathrm{t} = V_0 \times \mathrm{exp}(\frac{t}{4.5/\ln(V_0)})
\end{align}
where $t$ denotes the time in the future in billions of years. Using the same parameters as the simulations shown in Figure \ref{fig:timestack}, this shortened the future lifespan of the complex terrestrial biosphere by 100 million years at most, significantly less than the effect of strong seafloor weathering. 

\begin{figure}[ht]
    \centering
    \textbf{Impact of Seafloor Weathering Parameters on Future Lifespan of the 
    Terrestrial Biosphere At Different $\alpha$}
    \includegraphics[width=1\linewidth]{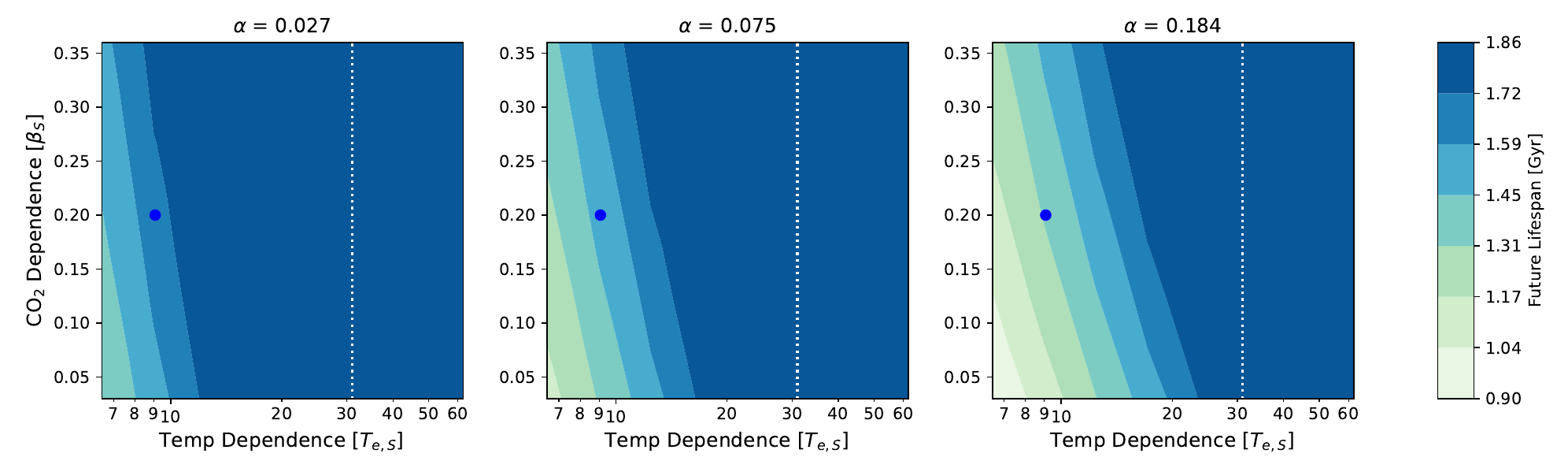}
    \caption{\textbf{The future lifespan of the terrestrial biosphere is only significantly shortened if seafloor weathering accounts for a large portion of global silicate weathering (large $\alpha$) and has a strong feedback (small $T_\mathrm{e,S}$).}
    Contour plots of the future lifespan of the terrestrial biosphere as a function of the temperature (${T\mathrm{_{e,S}}}$) and CO$_2$ ($\beta_{\mathrm{S}}$) dependencies of seafloor weathering. The dark points denote the median values of $T_\mathrm{e,S}$ and $\beta_\mathrm{S}$ from \citet{krissansen2017constraining}, and the middle panel corresponds to the median value of $\alpha$. The white dotted lines denote the temperature dependence of continental weathering (${T\mathrm{_{e,L}}}$).} 
    \label{fig:seafloorcontour}
\end{figure}

Above we found that the inclusion of a strong seafloor weathering feedback can shorten the future lifespan of the complex terrestrial biosphere by 300 million years (Fig.~\ref{fig:timestack}). However, the parameters governing seafloor weathering are poorly constrained \citep{krissansen2017constraining},and its effect on the future lifespan of the terrestrial biosphere varies over their plausible range (Fig.~\ref{fig:seafloorcontour}). 
We find that $T_{\mathrm{e,S}}$ and $\alpha$ can strongly impact the future lifespan, while $\beta_{\mathrm{S}}$ has a weaker effect. With the median values of $T_{e,L}$ and $\beta_L$ used here, for $T_{e,S}$ greater than 20 K, the long future lifespan found by Graham et al. (2024) comfortably holds for almost any value of $\alpha$ and $\beta_S$. Otherwise for small $T_{e,S}$, the impact of seafloor weathering on the future lifespan grows with the value of $\alpha$. The $T_{e,S}$ threshold dividing long and short future lifespans, however, is subject to the choice of $T_{e,L}$ and $\beta_L$, which we do not vary in this study. In order to determine whether seafloor weathering will impact the future lifespan of the complex terrestrial biosphere, therefore, $T_{\mathrm{e,S}}$ and $\alpha$ must be better constrained.

\section{Discussion}

Overall, consistent with \citet{graham2024substantial} and despite the impacts of seafloor weathering and stochastic variations in CO$_2$ outgassing, we find that Earth's terrestrial biosphere could persist for hundreds of millions of years longer than the $\sim$1 Gyr suggested in most previous studies \citep[e.g.,][]{caldeira1992life, lenton2001biotic, franck2002long, VonBloh2003, franck2006causes, rushby2018long, SousaMello2020, ozaki2021future, mello2023planetary}. For simulations with default parameter choices from \citet{graham2024substantial} based on estimates from \citet{krissansen2017constraining}, the terrestrial biosphere lifespan remains longer than $1.5$ Gyr unless modern seafloor weathering is larger than expected relative to continental weathering ($\alpha \gtrapprox 20\%$) and displays a significantly stronger temperature dependence ($T_\mathrm{e,S}<10$ K vs. $T_\mathrm{e,L}\approx40$ K). Similarly, unless the true magnitude of CO$_2$ outgassing variability is significantly larger than $10^{-4}$ Tmol Year$^{-1}$, stochastic variations in outgassing are unlikely to significantly hasten terrestrial extinction.

On Earth-like exoplanets there may be large variations in parameters like land fraction and seafloor spreading rate, such that the relative importance of continental vs. seafloor weathering covers the full range of possibilities from full land dominance to full sea dominance \citep[e.g.,][]{chambers2020effect,hayworth2020waterworlds}. Although we found that seafloor weathering has a relatively minor impact over a parameter regime that seems plausible for Earth, it would have a much larger effect on Earth's future trajectory if it made up a large fraction of modern planetary weathering. Variations in the balance between continental and seafloor weathering across planetary systems could therefore produce a large spread in CO$_2$ and climate for exoplanets subject to otherwise equivalent boundary conditions.

The future lifespan of Earth's complex terrestrial biosphere is a key input for at least two statistical models that attempt to leverage timings of events throughout Earth history to constrain whether the origin of life on Earth was a high-probability or low-probability occurrence. The ``Carter model'' \citep{carter1983anthropic,hanson1998must, carter2008five} posits that intelligent life emerged as the result of a series of one or more extremely unlikely evolutionary transitions or ``hard steps'' across Earth history. Although the validity of the model has been questioned on a variety of grounds \citep{vukotic2007timescale,simpson2017longevity,mills2025reassessment,whitmire2025abiogenesis}, it has seen significant use in the astrobiology literature \citep[e.g.,][]{hanson1998must,watson2008implications,waltham2017star, lingam2019role,hanson2021if,snyder2021timing,snyder2022catastrophe}. In the Carter model, a longer future biosphere lifespan leads to a greater likelihood that the origin of life was not an intrinsically unlikely ``hard step'' \citep{carter2008five}, and \citet{graham2024substantial} showed that with a future lifespan of 1.86 Gyr, the origin of life could be statistically ruled out as a hard step if origin of life timing were constrained precisely enough, whereas a future lifespan of 1 Gyr would prevent the ruling out of an unlikely origin of life. Similarly, the more recent ``Kipping model'' \citep{kipping2020objective} uses the future lifespan of the biosphere as an input for a Bayesian analysis of the probability of the origin of life, and, in line with the Carter model, a longer future lifespan leads to a larger probability that the origin of life was easy. If the origin of life is taken to have occurred at 4.1 Ga based on potentially biogenic carbon found in ancient zircons \citep{bell2015potentially}, then the 1.86 Gyr maximum future biosphere lifespan found in \citet{graham2024substantial} and this study produces a $\sim$10:1 odds ratio that the origin of life was an easy process, which constitutes ``strong evidence'' for easy abiogenesis in Bayesian terminology, whereas a 1 Gyr lifespan would not reach that evidence threshold \citep{kipping2025strong}. So, from the perspectives of the the Carter and Kipping models, seafloor weathering and modestly varying CO$_2$ outgassing seem unlikely to shorten the future lifespan of Earth's biosphere enough to preclude the possibility that the origin of life is an easy process on Earth-like planets. 

An obstacle to quantifying the impact of seafloor weathering on the future lifespan of Earth's terrestrial biosphere lies in its poorly constrained parameters. Of the three considered in this study ($T_\mathrm{e,S}$, $\beta_\mathrm{S}$, and $\alpha$), tighter constraints \citep[e.g.,][]{coogan2013evidence,krissansen2017constraining} on $\alpha$ 
would be an important target. If seafloor weathering turns out to be a small fraction of all modern silicate weathering, then barring a reorganization of the carbon cycle, regardless of the values of $T_\mathrm{e,S}$ and $\beta_\mathrm{S}$, seafloor weathering would be insignificant in the determination of the future lifespan of the terrestrial biosphere. 

There are a number of large uncertainties associated with the modeling done here. For example, if outgassing variation is stronger in the future than suggested by the Phanerozoic record \citep{baum2022simple}, premature CO$_2$ starvation or overheating in the near future could also be much more likely \citep[e.g.,][]{coy2025the}. Additionally, the OU model may fail to accurately represent the physical properties that govern outgassing rates when variability is high (e.g. producing periods of zero outgassing), and the model may have to be modified in such cases. The weathering and climate models have limitations as well. Variations in seafloor spreading rates, which impact the proportion of surface areas of land and sea exposed to silicate weathering \citep{krissansen2017constraining} and may couple with the global CO$_2$ outgassing rate \citep[e.g.,][]{franck1999modelling}, are not considered in our model. Likewise, we ignore potential variations in planetary albedo from changes in sea ice and cloud fraction, which could amplify climate sensitivity. Additionally, as we have emphasized throughout this study, the parameters that control the weathering response on land and the seafloor are poorly constrained. While we only examine parameter regimes in which seafloor weathering is a negative feedback, thus favoring death by CO$_2$ starvation, some research has proposed that silicate weathering could act as a positive feedback under warmer climate regimes due to the temperature dependence of thermodynamic equilibrium constants that govern coupled reactions of silicate dissolution and clay precipitation \citep{winnick2018relationships, hakim2021lithologic}. If Earth were to transition into a regime where silicate weathering acts as a positive feedback with respect to temperature, this could cause the carbon cycle equilibrium point where CO$_2$ drawdown by weathering matches CO$_2$ outgassing to change from stable to unstable \citep[e.g.,][]{Strogatz-1994:nonlinear}, such that the coupled climate-carbon cycle system would tend to amplify perturbations instead of suppressing them. This would likely lead to rapid destabilization of the carbon cycle, with CO$_2$ being pushed to uninhabitably high or low levels within a few million years at most \citep{broecker1998does,zeebe2008close,colbourn2015time}. 

\section{Conclusion}

In this study, we applied a global-mean model of Earth's climate and carbonate-silicate cycle, incorporating seafloor weathering and stochastic outgassing to reevaluate the lifespan of the complex terrestrial biosphere under a brightening Sun. We showed that across a wide range of seafloor weathering assumptions, a future lifespan of $\sim$1.86 Gyr, as suggested by \citet{graham2024substantial}, remains plausible, and a lifespan longer than $\approx$1 Gyr is very likely. The largest reductions in the future lifespan of the complex terrestrial biosphere occur in cases where seafloor weathering is a high fraction of total weathering and the seafloor weathering feedback is much stronger than the continental weathering feedback. Interestingly, in these cases the kill mechanism switches from overheating to CO$_2$ starvation. For stochastic outgassing parameters that are consistent with the reconstructions of outgassing from the past 1 billion years, the probability of shortening the future lifespan of the terrestrial biosphere so dramatically such that it falls below 1 billion years remains below 50\%.

\section{Acknowledgments}

This work was partially supported by a University of Chicago Quad Faculty Research Grant.

Code to reproduce the figures in this article is available in a Zenodo repository \citep{zhu_scripts_2025}.

%

\vspace{5mm}

\software{Matplotlib \citep{hunter2007matplotlib}, Numpy \citep{harris2020array}, Scipy \citep{jones2001scipy}}

\appendix

\bibliography{biblio,biblio2}{}
\bibliographystyle{aasjournal}



\end{document}